\documentclass[10pt,showkeys]{revtex4} \textwidth 17cm \textheight 24cm
\usepackage{psfig,epsfig}
\newcommand{\lr}[1]{\langle #1 \rangle}
\newcommand{\pcf}[2]{{\cal D}_{#1}(#2)}
\newcommand{\bi}[1]{Fig.~\ref{fig:#1}}
\newcommand{\e}[1]{eq.~(\ref{#1})}
\begin{document}
\title{\bf Moments of the first passage time under external driving}
\date{\today}
\author{Benjamin Lindner} 
\affiliation{Department of Physics, University of Ottawa, 150 Louis Pasteur,
Ottawa, Canada KIN 6N5}
\begin{abstract}
  A general theory is derived for the moments of the first passage
  time of a one-dimensional Markov process in presence of a weak
  time-dependent forcing. The linear corrections to the moments can be
  expressed by quadratures of the potential and of the time-dependent
  probability density of the unperturbed system or equivalently by its
  Laplace transform. If none of the latter functions is known, the
  derived formulas may still be useful for specific cases including a
  slow driving or a driving with power at only small or large times.
  In the second part of the paper, explicite expressions for mean and
  variance of the first passage time are derived for the cases of a
  linear or a parabolic potential and an exponentially decaying
  driving force.  The analytical results are found to be in excellent
  agreement with computer simulations of the respective first-passage
  processes. The particular examples furthermore demonstrate that
  already the effect of a simple exponential driving can be fairly
  involved implying a nontrivial nonmonotonous behavior of mean and
  variance as functions of the driving's time scale.
\end{abstract}
\pagestyle{myheadings}
\markboth{Benjamin Lindner}{First passage times under external driving}
\setcounter{page}{1}
\maketitle
\section{Introduction}
One of the key results in the theory of stochastic processes are the
quadrature expressions for the moments of the first passage time (FPT)
in case of a one-dimensional Markov process \cite{PonAnd33,Sie51}.
One of the assumptions made in the seminal papers by Pontryagin et
al. \cite{PonAnd33} and Siegert \cite{Sie51} is the temporal
homogeneity of the process: except for the driving Gaussian white
noise the system is subject to a force without time dependence.  In
many instances this assumption does not hold and an extension of the
classic theory to the case of a time-dependent force is desirable.\\
In the context of many noise-induced phenomena, for instance, the
presence of an additional deterministic or stochastic perturbation is
essential and its effect on the passage time statistics is of foremost
interest. In case of {\em resonant activation} (RA)
\cite{DoeGad92,BieAst93,PecHan94,Rei95}, the additional driving is a
stochastic process with state-dependent amplitude.  For the problems
of {\em stochastic resonance} (SR) (see Ref.~\cite{GamHan98} and
references therein) and {\em coherent stochastic resonance} (CSR)
\cite{FleHav88,MasRob95,GitWei95,Por97} the driving is commonly a
deterministic periodic signal. The key feature of both RA and CSR is a
minimum in the mean FPT or mean exit time attained at a finite
``optimal'' value of the forcing's time scale (e.g. correlation time
or driving period). Similarly, SR is realized if a time-scale matching
condition between the forcing period and the FPT of the unperturbed
system (an inverse Kramers rate) is met.\\
My own interest in the general problem originates in studies of
stochastic neuron models involving an exponentially decaying
time-dependent forcing. For these models, the FPT corresponds to the
interspike intervals (ISIs) separating the neural discharges (action
potentials or spikes) by means of which neurons transmit and process
signals. The exponentially decaying perturbation in these models
arises from slow ionic currents that are driven by the spiking
activity of the neuron itself \cite{LinLon03b}. An exponential
perturbation is also obtained via a simple transformation of models
with a decaying threshold; such models have been employed to reproduce
the statistics of certain sensory neurons \cite{ChaLon00,ChaPak03}. In
general, the effect of an exponentially decaying driving on the FPT
statistics is poorly understood up to now in contrast to the
frequently studied case of periodically forced stochastic neuron
models (see
e.g. Refs.~\cite{Lon93,BulLow94,BulEls96,PleTan97,Lan97,PleGei01}).\\
In order to study the change of the FPT statistics induced by a
time-dependent driving, researchers have used two different analytical
approaches. First, keeping the potential shape, the driving force as
well as the boundary and initial conditions as simple as possible
allows in some cases for an exact, in others for an approximate
solution. This kind of approach was pursued in the late 1980's
\cite{FleHav88,BalVan88,Bez89} and during the 1990's
\cite{DoeGad92,BieAst93,BulLow94,GitWei95,Por97} (see also
Ref.~\cite{Red01}, ch.~4 for further examples); in most cases
(piecewise) linear potentials and a periodic or dichotomous driving
were considered. Later several researchers proposed semi-analytic
procedures to solve these FPT problems
\cite{Men92,KlikYao01,ChoFox02}. It should also be noted that there exists
a considerable mathematical literature on the FPT problem with
time-dependent deterministic forcing, most of which is devoted to
specific simple cases that are exactly solvable by a transformation to
a time-homogeneous system (see Refs.~\cite{GutGon91,GutRom95,GutRic97}
and further references cited in these papers).\\
The other analytical approach consists of a weak noise analysis of
driven first-passage processes. Here the only assumption made about
the potential is the existence of a metastable state.
Refs.~\cite{SmeDyk99,LehRei00,Shu00,TalLuc03} focused on the escape
rate out of a metastable one-dimensional potential (the inverse mean
first passage time for a quasi-equilibrium initial condition) in the
small-noise limit and aimed thus at a generalization of the famous
Kramers problem \cite{Kra40,HanTal90} for periodic
\cite{SmeDyk99,LehRei00,TalLuc03} or stochastic
\cite{PecHan94,Rei95} driving. Although both approaches  led to 
significant progress, there are also many instances where one wishes
to go beyond the limit set by assuming weak noise or piecewise linear
potentials and where also higher moments and not only the mean FPT (or
its inverse) may be of interest.\\
Here I give an extension of the classic concept due to Pontryagin
\cite{PonAnd33} and Siegert \cite{Sie51} of calculating the FPT
moments for a general potential to the case that includes a weak
time-dependent driving. I focus on a deterministic driving function;
the approach might be, however, also helpful in situations with an
additional stochastic driving.  I will use a perturbation calculation
that requires a weak driving and will consider the first (linear)
correction term for each moment. No restrictions apply with respect to the
noise intensity or the time scale of the driving, although the range
of validity of the theory, i.e. the question how ``weak'' the driving
has to be, may also depend on these parameters. My most general result
relates the linear corrections to each moment of the FPT to
quadratures of the time-dependent probability density $P_0(x,t)$ of
the unperturbed system or to quadratures of the Laplace transform of
this function. An alternative formulation provides the corrections in
terms of an infinite sum of quadratures of known functions. These
results cannot be applied in the general case (arbitrary driving
function and arbitrary potential shape), since $P_0(x,t)$ and/or its
Fourier transform are not known for most potentials and the numerical
evaluation of an infinite sum of quadratures is in general not
feasible either. For many important cases, however, including a slow
driving, a driving with power at only small times or a driving with
power at only large times, the derived general formulas can be useful
for the FPT problem in a potential of arbitrary shape. Furthermore, I
will show that for two specific potential shapes (linear and
parabolic) and an exponentially decaying driving force, the derived
theory yields valid (and in part strikingly simple) results for the
FPT's mean and variance for arbitrary time scale of the driving (slow,
moderate, or fast compared to the mean FPT of the unperturbed
system). The explicite results derived for an exponential driving
function can be applied to the neurobiological problems mentioned
above; this will be done elsewhere. Since the exponential decay of
drift parameters is frequently encountered in many situations, the
results may be also helpful for the study of other physical systems.\\
This paper is organized as follows. Sec.~\ref{sec:general} starts with
an introduction of the problem and of the quantities of interest. Next
I present an derivation of alternative quadrature formulas for the FPT
moments in the time-homogeneous case. These quadratures are entirely
equivalent to the classic formulas by Siegert \cite{Sie51} as shown in
the appendix. The alternative approach will be the basis for the
perturbation calculation of the time-inhomogeneous first-passage
problem leading to the general relation between the corrections of the
moments and the quadratures of the unperturbed probability density.
In sec.~\ref{sec:special}, I derive explicite results for a linear and
a parabolic potential and an exponential driving force. These
analytical results will be compared to simulations of the two systems
in sec.~\ref{sec:simus}. Here I will show that, remarkably, similar to
the case of periodic driving, a nontrivial behavior of the FPT's mean
and variance with respect to the time scale of the driving (i.e. the decay rate
of the exponential driving) is possible. The mechanisms for these
``resonances'' will be discussed.  Sec.~\ref{sec:summary} summarizes
the findings and discusses further applications and extensions of the
theory.\\
\section{General Theory} 
\label{sec:general}
\subsection{Langevin and Fokker-Planck equations} 
Starting point of my consideration is the Langevin equation for a
one-dimensional escape process given by a potential $U(x)$, a
white-noise driving of intensity $D$, and an additional weak
time-dependent forcing $\varepsilon s(t)$:
\begin{equation}
\label{langevin}
\dot{x}=-U'(x)+\varepsilon s(t)+\sqrt{2D} \xi(t)
\end{equation}
Without loss of generality I consider an initial value at zero
($x(0)=0$) and ask for the first-passage time to a point $x_E$ to the
right of the origin ($x_E>0$). The only restriction for the potential
is that I exclude potentials that allow for an escape toward minus
infinity. Furthermore, I do not specify the forcing $s(t)$ but take
for granted that its effect on the FPT statistics is weak by virtue of
the small parameter $\varepsilon$.\\
Instead of using the Kolmogorov (backward) equation as it is commonly
done in the treatment of FPT problems \cite{Sie51,Gar85}, I shall
employ the Fokker-Planck (forward) equation (FPE) governing the
probability density of $x$
\begin{equation}
\label{fpe}
\partial_t P(x,t)=\partial_x[ ( U'(x)-\varepsilon s(t)) P(x,t)+D\partial_x P(x,t)].
\end{equation}
The FPT problem stated above determines the initial condition (at
$t=0$ the variable $x$ is with certainty at $x=0$) and the boundary
condition (absorbing boundary condition at $x=x_E$)
\begin{eqnarray}
P(x,0)&=&\delta(x), \\ 
P(x_E,t)&=& 0.
\end{eqnarray}
It is well known that there is a simple relation between the FPT
density and a quantity derived from the probability density $P(x,t)$:
the FPT density is given by the time-dependent probability current at
the absorbing boundary (see, for instance, Ref.~\cite{Hol76})
\begin{eqnarray}
\label{fptd_j} 
\Phi(T)&=&J(x_E,T)=-[ ( U'(x)-\varepsilon s(T)) P(x,T)+D\partial_x P(x,T)]|_{x=x_E} \nonumber \\
&=&-D\partial_x P(x,T)|_{x=x_E}. 
\end{eqnarray}
Provided  $\Phi(t)$ is known, one can calculate the moments of the FPT by
\begin{equation} 
\label{moments_def} 
\langle T^n(0 \to x_E)\rangle= \int\limits_0^\infty dT\; T^n \Phi(T)= -D \int\limits_0^\infty dT\; T^n \partial_x P(x,T)|_{x=x_E}. 
\end{equation}
For certain problems it may be desirable to know also the Laplace
transform of $\Phi(T)$. This function can be expressed by the Laplace
transform of the probability density $\tilde{p}(x,p)$ as follows
\begin{equation}
\label{rho_def}
\rho(p)=\int\limits_0^\infty dt\; e^{-p t} \Phi(t)=-D\partial_x \tilde{p}(x,p)\Bigg|_{x=x_E}.
\end{equation}
By means of the Laplace transform the moments can be calculated as follows
\begin{equation} 
\label{moments_rho} 
\langle T^n(0 \to x_E)\rangle= (-1)^n \frac{d^{(n)}}{dp^n} \rho(p)\Bigg|_{p=0}.
\end{equation}
However, even in the absence of a time-dependent driving ($\varepsilon
= 0$), solving directly for one of the functions $P(x,t)$, $\Phi(T)$,
or $\rho(p)$ is possible only in a few simple cases including constant
and linear potential shapes. Nevertheless, for calculating the moments
of the FPT we are not restricted to use \e{moments_def} or
\e{moments_rho}. Remarkably if $\varepsilon=0$, the moments of the FPT {\em can } 
be calculated {\em exactly} for an {\em arbitrary} potential shape
$U(x)$. The general formula for the $n$-th moment of the passage from
a general initial point $x=a$ to an absorbing boundary at $x=b$ is
given by \cite{Sie51}
\begin{equation} 
\label{quadratures} 
\langle T^n(a\to b)\rangle_0= \frac{n}{D} \int_a^{b} dy e^{U(y)/D} 
\int_{-\infty}^y dx e^{-U(x)/D}  \langle T^{n-1}(x\to b) \rangle_0  
\end{equation}
where the index ``0'' indicates that $\varepsilon=0$ (later on, this
convention will be also applied to the functions $P(x,t),
\tilde{p}(x,p)$ and $\rho(p)$).  In order to calculate the $n$-th
moment one has to solve for the lower moments as functions of the
initial point first. The hierarchy of quadratures is completed by
stating that for obvious reasons $\langle T^0(a\to b)\rangle= \langle
1 \rangle=1$.\\
The aim of this paper is to extend these expressions to the case of a
weak time-dependent driving function $\varepsilon s(t)$. In other
words, I seek for the linear correction term to the n-th moment
denoted $j_n(x_E)$ such that
\begin{equation}
\lr{T^n}=\lr{T^n}_0+\varepsilon j_n(x_E).
\end{equation} 
In particular, once the corrections to the first two moments have
been calculated, mean and variance of the FPT will be to linear order
in $\varepsilon$ given by
\begin{eqnarray}
\lr{T}&=&\lr{T}_0+\varepsilon j_1(x_E),\\
\lr{\Delta T^2}&=& \lr{T^2}-\lr{T}^2\nonumber \\
&=&\lr{\Delta T^2}_0+\varepsilon [j_2(x_E)-2\lr{T}_0 j_1(x_E)].
\end{eqnarray}
Later on, for specific systems, I will solely discuss these first two
cumulants and the relative standard deviation (coefficient of
variation) of the FPT, that is a function of them
\begin{equation}
\label{cv_def}
CV=\frac{\sqrt{\lr{\Delta T^2}}}{\lr{T}}.
\end{equation}
\subsection{Moments of the first passage time in the autonomous case -
the other way around}
Here I set $\varepsilon=0$. First I introduce the functions
\begin{equation}
\label{beta_def}
\beta_n(x)=\int_0^\infty dt \;\; t^n P(x,t) 
\end{equation}
and
\begin{equation}
\label{jn_def}
J_n(x)=-(U'(x)+D\frac{d}{dx}) \beta_n(x) 
\end{equation}
On comparing \e{moments_def} and eqs.~(\ref{beta_def}) and
(\ref{jn_def}) it becomes apparent that $\beta_n$ and $J_n$ are
related to the $n$-th moment of the FPT as follows
\begin{equation}
\label{rel_Tn_Jn}
\langle T^n \rangle_0= -D \beta'_n(x_E)=J_n(x_E).
\end{equation}
Here and in the following the prime denotes the derivative with respect to $x$.
I now derive the general solutions for $J_n(x)$ that provides an
alternative solution for the FPT moments by virtue of \e{rel_Tn_Jn}.\\
Multiplying the FPE (\ref{fpe}) with $t^n$, integrating
over $t$, and using integration by part on the l.h.s. of
the equation, I obtain for the functions $\beta_n(x)$ 
the following set of equations
\begin{eqnarray} 
-\delta(x)&=&\frac{d}{dx}(U'(x)+D \frac{d}{dx}) \beta_0(x), \\
-n \beta_{n-1}(x) &=&\frac{d}{dx}(U'(x)+D \frac{d}{dx}) \beta_n(x), 
\end{eqnarray}
The boundary conditions can be inferred from those for $P_0(x,t)$ 
\begin{equation}
\label{bc_beta}
\beta_n(x_E)=0 \;\;\;\;\; \mbox{and} \;\;\;\; \beta_n^{(k)}(-\infty)=0\;\; (\mbox{with} \;\; k=0,1,2,\cdots)
\end{equation}
with $\beta_n^{(k)}(x)$ denotes the $k$-th derivative.
The solutions are straightforward
\begin{eqnarray}
\label{beta_0}
\beta_0(x)&=&\frac{1}{D}e^{-U(x)/D}\int\limits_x^{x_E} dy e^{U(y)/D} \Theta(y),\\
\label{beta_n}
\beta_n(x)&=&\frac{n}{D}e^{-U(x)/D}\int\limits_x^{x_E} dy e^{U(y)/D} \int\limits_{-\infty}^y dz\; \beta_{n-1}(z), \;\;\;\;\;\;\;\;\; n=1,2,\cdots
\end{eqnarray}
where $\Theta(x)$ denotes the Heaviside jump function \cite{AbrSte70}.
For the $J_n(x)$, I find 
\begin{eqnarray} 
\label{hierar_Jn_beta_0}
-\delta(x)&=&- J_0'(x), \\
\label{hierar_Jn_beta_n}
-n \beta_{n-1}(x) &=&- J_n'(x) , \;\;\;\;\;\;\;\;\; n=1,2,\cdots
\end{eqnarray}
The first equation leads immediately to a Heaviside function 
\begin{equation}
\label{j0_0}
J_0(x)=\Theta(x),
\end{equation}
To the second equation I apply the operator $-(U'(x)+Dd/dx)$
which yields
\begin{equation}
\label{hierarchy}
-n J_{n-1}(x)= U'(x) J'_n(x)+D J''_n(x), \;\;\;\;\;\;\;\;\; n=1,2,\cdots
\end{equation}
From eqs.~(\ref{hierar_Jn_beta_0}),(\ref{hierar_Jn_beta_n}), and
(\ref{bc_beta}), I get boundary conditions for the $J_n$
\begin{equation}
J'_n(x_E)=0,\;\;\; J_n(-\infty)=0
\end{equation}
The solution of \e{hierarchy} obeying these conditions reads 
\begin{equation}
\label{jn_0}
J_n(x)=\frac{n}{D} \int\limits_{-\infty}^x dy\; e^{-U(y)/D} \int\limits_y^{x_E} dz\; e^{U(z)/D} J_{n-1}(z)
\end{equation}
This equation together with \e{j0_0} and \e{rel_Tn_Jn} provides an
alternative, though not in the least simpler way of calculating the
moments of the FPT. On comparing with \e{quadratures}, I note the
differences in the signs of the exponents, in the boundaries of
integration, and in the first function of the hierarchy
\e{j0_0} (for \e{quadratures}, the first of the functions was $\lr{T^0(a\to
b)}=1$). Nevertheless, the alternative quadrature expressions derived
here are completely equivalent to \e{quadratures} as shown in the
appendix. In particular for $n=1$ \e{jn_0} and \e{j0_0} yield
\begin{eqnarray}
\lr{T}_0=J_1(x_E)&=&\frac{1}{D}\left[ \int_{-\infty}^0 dx\;\; e^{-U(x)/D} \int_0^{x_E} dy\;\; e^{U(y)/D}+ \right.\nonumber \\
&& \hspace{2em} \left. \int_{0}^{x_E} dx\;\; e^{-U(x)/D} \int_x^{x_E} dy\;\; e^{U(y)/D} \right] 
\end{eqnarray}
Changing the order of integration yields then the same result as
eq.~(\ref{quadratures}) for $n=1, a=0, b=x_E$.
\subsection{Including a weak time-dependent force} 
I now turn to the case $\varepsilon\neq 0$. For the probability density 
$P(x,t)$ obeying the FPE (\ref{fpe}), I make the following ansatz
\begin{equation}
P(x,t)=P_0(x,t)+\varepsilon q(x,t)
\end{equation}
where the first term is the probability density for $\varepsilon=0$.
Omitting all terms of order $\varepsilon^2$ and higher, I find the following
equation governing the second function
\begin{equation}
\label{fpe_q}
\partial_t q(x,t)=\partial_x(U'(x)+D\partial_x)q(x,t) - s(t) \partial_x P_0(x,t)
\end{equation}
The boundary and initial  conditions for this function can be inferred from those 
of $P(x,t)$ and $P_0(x,t)$
\begin{eqnarray}
\label{bc_q_1} P(x_E,t)\equiv P_0(x_E,t)\equiv 0 \;\;\;\; & \rightarrow & \;\;\; q(x_E,t)\equiv 0 \\
\label{bc_q_2} P(x,0)=P_0(x,0)=\delta(x) \;\;\;  & \rightarrow & \;\;\; q(x,0)\equiv 0 \\
\label{bc_q_3} P(-\infty,t)\equiv P_0(-\infty,t) \equiv 0 \;\;\;\;\;\;\; & \rightarrow & \;\;\; q(-\infty,t)\equiv 0    
\end{eqnarray}
Now I introduce the counterpart to the functions $J_n(x)$ corresponding to
the perturbation $q(x,t)$
\begin{equation}
j_n(x)=-(U'+D\partial_x) \int_0^\infty dt \;\; t^n q(x,t)
\end{equation}
Knowledge of this function allows for the calculation of the $n$-th moment
of the first passage time by the following formula
\begin{equation}
\lr{T^n}=\lr{T^n}_0+\varepsilon j_n(x_E).
\end{equation}
From \e{fpe_q}, I find 
\begin{equation}
\label{jn_p0}
-j'_n(x)=-n\int_0^\infty dt \;\; t^{n-1} q(x,t)+\int_0^\infty dt \;\;t^n s(t) \partial_x P_0(x,t)
\end{equation}
From this equation and from \e{bc_q_3} and \e{bc_q_1}, I can conclude that 
\begin{eqnarray}
\label{bc_jn_1}
&& j_n(-\infty)=0 \\
\label{bc_jn_2}
&& j'_n(x_E)=-\int_0^\infty dt \;\;t^n s(t) \partial_x P_0(x_E,t)=-\partial_x \hat{\cal T}_n P_0(x,t)
\end{eqnarray} 
where I have used the integral operator $\hat{\cal T}_n$ that is defined by 
\begin{equation}
\hat{\cal T}_n =\int_0^\infty dt\;\; s(t) t^n 
\end{equation}
(function is multiplied by $t^n s(t)$ and then integrated).
The solution for $j_0(x)$ is straightforward
\begin{equation}
\label{j0}
j_0(x)=-\hat{\cal T}_0  P_0(x,t)=-\int_0^\infty dt\;\; s(t) P_0(x,t)
\end{equation}
The constant of integration in \e{j0} is zero because of
the boundary conditions \e{bc_q_1} and \e{bc_jn_1}.\\
For $n>0$, I apply the operator $-(U'+D\partial_x)$ to \e{jn_p0}
\begin{equation}
U' j'_n + D j''_n = -n j_{n-1} - \hat{\cal T}_n (U'+D\partial_x) P_0(x,t)=f_n(x)
\end{equation}
This equation has the same structure like those for the functions
$J_n(x)$ of the unperturbed system. The difference lays in the
inhomogeneities on the r.h.s. (abbreviated by $f_n(x)$) and the
different boundary condition for the derivative of $j_n$ at $x_E$.\\
The solution for the derivative is given by 
\begin{eqnarray}
j'_n(x)&=& e^{-\textstyle\frac{U(x)}{D}}\left[c_n+\frac{1}{D}\int\limits_{-\infty}^x dy\;\; e^{\textstyle\frac{U(x)}{D}} f_n(y) \right] \nonumber \\
&=& e^{-\textstyle\frac{U(x)}{D}}\left[c_n-\frac{n}{D}\int\limits_{-\infty}^x dy\;\; e^{\textstyle\frac{U(x)}{D}} j_{n-1}(y) -\hat{\cal T}_n e^{\textstyle\frac{U(x)}{D}} \partial_y P_0(y,t)\right|_{-\infty}^{x} \!\!\!\!+\nonumber  \\
&& + \left. \hat{\cal T}_n\left(\int\limits_{-\infty}^x dy\;\; e^{\textstyle\frac{U(x)}{D}} \partial_y^2 P_0(y,t)-\int\limits_{-\infty}^x dy \;\; e^{\textstyle\frac{U(x)}{D}} \partial_y^2 P_0(y,t) \right) \right]
\end{eqnarray}
The terms in the last line cancel and the integration constant $c_n$
has to be chosen such that condition \e{bc_jn_2} is met. I
obtain
\begin{equation}
j'_n(x)=-\hat{\cal T}_n  \partial_x P_0(x,t)+\frac{n}{D}e^{-U(x)/D} \int_{x}^{x_E} dy\;\; e^{U(y)/D} j_{n-1}(y)
\end{equation}
Another integration (for which  the integration constant is determined by the boundary condition
at $-\infty$ given in \e{bc_jn_1}) yields 
\begin{equation}
\label{jn}
j_n(x)=-\hat{\cal T}_n  P_0(x,t)+\frac{n}{D} \int_{-\infty}^{x} dz\;\; e^{-U(z)/D} \int_{z}^{x_E} dy\;\; e^{U(y)/D} j_{n-1}(y)
\end{equation}
This equation is the first important result of my paper.  I recall
that the corrections to the moments of the first passage time are
obtained by taking $j_n$ at $x_E$. Thus {\em if the function
$P_0(x,t)$ is given}, one may calculate the effect of the external
driving on an arbitrary moment of the first passage time by a
subsequent solution of all the lower moments. If $P_0(x,t)$ is {\em not}
given (which is, unfortunately, usually the case) there are still 
several classes of tractable problems for which \e{jn} is useful.
These are discussed in the next subsection.
\subsection{Further simplification of the result for specific cases}
Eq.~(\ref{jn}) involves integrals over the probability density of the
unperturbed system multiplied with the time-dependent part of the
drift and powers of $t$. Expanding the driving function in powers of
$t$ permits to express these integrals in terms of the known functions
$\beta_k$ from \e{beta_n} as follows
\begin{equation}
\hat{\cal T}_n P_0(x,t) =\int_0^\infty dt\;\; s(t) t^n P_0(x,t)=\sum\limits_{k=0}^\infty \frac{s^{(k)}(0)}{k!} \beta_{k+n}(x) 
\end{equation}
by means of  which I obtain 
\begin{equation}
\label{jn_beta}
j_n(x)=\frac{n}{D} \int_{-\infty}^{x} dz\;\; e^{-U(z)/D} \int_{z}^{x_E} dy\;\; e^{U(y)/D} j_{n-1}(y) -\sum\limits_{k=0}^\infty \frac{s^{(k)}(0)}{k!} \beta_{k+n}(x)
\end{equation}
This formula is especially useful in case of a slow driving that can
be for $t\sim \lr{T}_0$ described by just a few expansion
terms $s^{(k)}$. As can be expected, the zeroth term results in the
static correction, what I briefly show now for the simplest case
$n=1$. Suppose a static driving $s(t)=1$, then $j_0(x)=-\beta_0(x)$
and the linear correction to the mean FPT reads
\begin{eqnarray}
\label{j1_static}
j_{1,static}(x_E)&=& -\frac{1}{D} \int\limits_{-\infty}^{x_E} dz\; e^{-U(z)/D} \int\limits_{z}^{x_E} dy\; e^{U(y)/D} \beta_0(y)\nonumber \\
&=&-\frac{1}{D^2} \int\limits_{-\infty}^{x_E} dz\;\; e^{-U(z)/D} \int\limits_{z}^{x_E} dx \int\limits_x^{x_E} dy\; e^{U(y)} \Theta(y) \;\;  \nonumber \\
&=&\frac{1}{D} \int\limits_{0}^{x_E} dy\int\limits_{-\infty}^{y} dz\; \frac{z-y}{D} e^{[U(y)-U(z)]/D} 
\end{eqnarray}
This correction is also obtained by considering the unperturbed system
with a potential $\tilde{U}(x)=U(x)-\varepsilon x$, writing down the mean FPT
according to \e{quadratures} with $n=1, a=0, b=x_E$, and expanding the
result up to first order in $\varepsilon$. For a non-static but slow forcing,
the first correction term describing a truly dynamical effect of the
driving would be obtained by taking into account a finite $s'(0)$,
leading to a quadrature formula that involves $\beta_1(x)$. Although
the incorporation of higher non-static corrections is straightforward,
note that the number of quadratures which have to be numerically
solved increases by two with every term $s^{(n)}(0)$ that is taken
into account.\\
Besides a slow driving another important class of perturbations is
given by drivings described by an exponential decay or a periodic
function, both of which can be described by $\exp[-\lambda t]$ or a
sum of such exponentials. In this case the term $\hat{\cal T}_n
P_0(x,t)$ equals the n-th derivative of the Laplace transform
$\tilde{p}_0(x,p)$ of $P_0(x,t)$ with respect to the complex argument $p$ taken
at $p=\lambda$
\begin{eqnarray}
\label{TP_0_expdriv}
\hat{\cal T}_n P_0(x,t)&=&\int_0^\infty dt\;\;
e^{-p t} t^n P_0(x,t)\Bigg|_{p=\lambda}=(-1)^n \frac{d^n}{dp^n} \tilde{p}_0(x,p)\Bigg|_{p=\lambda} 
\end{eqnarray} 
Using this form in \e{jn} is in particular of advantage if the
function $\tilde{p}_0(x,p)$ is known but $P_0(x,t)$ is not known.  For
more general driving functions, this can be generalized as follows.
Suppose the Laplace transform of the driving $s(t)$ exists
\begin{equation}
\tilde{s}(\omega)=\int\limits_{0}^\infty dt\; e^{-p t} s(t),\;\;\;\; (p\ge0)
\end{equation}
Then it is possible to recast eqs.~(\ref{jn}) into the following form
\begin{equation}
\label{jn_om}
j_n(x)=-\hat{\cal F}_n \tilde{p}_0(x,-p)+\frac{n}{D}
\int_{-\infty}^{x} dz\;\; e^{-U(z)/D} \int_{z}^{x_E} dy\;\; e^{U(y)/D}
j_{n-1}(y)
\end{equation}
where the operator $\hat{\cal F}_n$ is defined by 
\begin{equation}
\hat{\cal F}_n = \frac{1}{2\pi i}\int\limits_{-i \infty}^{i \infty} dp \; \tilde{s}(p) \frac{d^n}{dp^n} 
\end{equation}
With a pure exponential driving, the term $\hat{\cal F}_n
\tilde{p}_0(x,-p)$ reduces to \e{TP_0_expdriv} as can be shown by the
calculus of residues.\\
Further simplifications or approximations are possible by means of
short-time or asymptotic approximations of $P_0(x,t)$ (see, for
instance, Ref. \cite{Van93}), if most of the driving power is at short
or long times. Here I shall not further dwell on the general case but
study the effect of a simple exponential driving on systems with
linear or parabolic potential for which exact correction formulas for mean
and variance can be found.
\section{Theory for  a system
with linear or parabolic potential and exponential driving}
\label{sec:special}
In the following, I will focus on the corrections to the first two
moments. These are given by
\begin{equation}
\label{j1}
j_1(x_E)=-\frac{1}{D}\int\limits_{-\infty}^{x_E} dx e^{-U(x)/D} \int_{x}^{x_E} dy\;\; e^{U(y)/D} \hat{\cal T}_0 P_0(y,t)
\end{equation}
\begin{eqnarray}
\label{j2}
j_2(x_E)&=&-\frac{2}{D}\int_{-\infty}^{x_E} dx\;\;e^{-U(x)/D} \int_{x}^{x_E} dy\;\; e^{U(y)/D} \left\{\hat{\cal T}_1 P_0(y,t) + \right.\nonumber\\
&& \left.\;\;\; \frac{1}{D} \int_{-\infty}^{y} dx_2 e^{-U(x_2)/D} \int_{x_2}^{x_E} dy_2\;\; e^{U(y_2)/D} \hat{\cal T}_0 P_0(y_2,t) \right\}
\end{eqnarray}
Furthermore, the following driving force is considered
\begin{equation} 
 s(t)=\lambda \exp[-\lambda t]. 
\end{equation}
Some remarks regarding this function are indicated.  The function
$s(t)$ is normalized (integration over time yields one).  There are
two simple limits: (i) for $\lambda \ll \lr{T_0}$, the function tends
to a static bias of amplitude $\lambda$; (ii) for $\lambda\to
\infty$ the function approaches a $\delta$ function that changes the
initial position from 0 to $\varepsilon$. In both limits, the moments
of the first passage time can be exactly calculated which gives us a
mean to check the validity of the perturbation
calculation. Furthermore, according to \e{TP_0_expdriv} the operators
$\hat{\cal T}_n$ correspond to derivatives of the Laplace transform
$\tilde{p}_0(x,\lambda)$ multiplied with $\lambda$
\begin{eqnarray}
\label{hat_TnP0}
\hat{\cal T}_n P_0(x,t)&=&\lambda (-1)^n \frac{d^n}{d\lambda^n} \tilde{p}_0(x,\lambda) 
\end{eqnarray} 
Finally note, that the correction formulas can be looked upon as
linear operations {\em on the driving function}. This implies that the
correction to a driving consisting of a sum of exponentials equals the
sum of the corrections to the single exponentials. In particular, this
applies to a (possibly damped) cosine driving $s(t)=\exp[-(\lambda
+i\omega)t]+c.c.$ with $\lambda,\omega\in \Re$ and $\lambda\ge 0$.\\
In the following, I will furthermore use a parabolic potential
\begin{equation}
\label{pot}
U(x)=b \frac{x^2}{2} -a x, (b\ge 0)
\end{equation} 
and will separately discuss the cases $b=0$ and $b\neq 0$. The former
problem corresponds with $a>0$ to a biased random walk toward the
absorbing boundary; there is no potential barrier present in this
simple case and the first passage will take place in a limited time
even at vanishing noise. In contrast to this, for $b\neq 0$ and
$a/b<x_E$ there exists a metastable point (potential minimum of
$U(x)$) to the left of the absorbing boundary; the first passage
process is noise-assisted, i.e. for vanishing noise the passage time
tends to infinity. I note that both cases are of particular importance
in the neurobiological context, where the FPT corresponds to the so
called interspike intervals generated by a perfect ($b=0$) or leaky
($b>0$) integrate-and-fire neuron stimulated by white noise
\cite{Hol76}.\\
Before I come to the specific cases, some further simplifications of
the correction formulas \e{j1} and \e{j2} for the potential
\e{pot} with arbitrary $b$ will be performed. It will emerge that for
a general parabolic potential the corrections can be entirely
expressed by the Laplace transform $\rho_0(\lambda)$ of the FPT
density of the unperturbed system instead by the function
$\tilde{p}_0(x,\lambda)$.\\
\subsection{Simplification of the correction formulas for arbitrary $b$}
Using \e{hat_TnP0} with $n=0$ the correction to the mean FPT \e{j1}
reads
\begin{equation}
\label{j1_2}
j_1(x_E)=-\frac{\lambda}{D}\int\limits_{-\infty}^{x_E} dx e^{-U(x)/D} \int_{x}^{x_E} dy\;\; e^{U(y)/D} \tilde{p}_0(y,\lambda).
\end{equation}
By multiplying the FPE \e{fpe} with $e^{-\lambda t}$ and integrating
over time, it is readily verified that $\tilde{p}_0(x,\lambda)$
appearing in \e{j1_2} obeys the following ordinary differential
equation
\begin{equation}
\label{fpe_lapla}
-\delta(x)+\lambda \tilde{p}_0=\frac{d}{dx}(U'+D\frac{d}{dx})\tilde{p}_0.
\end{equation}
Using this equation in the form
\begin{equation}
\label{fpe_lapla2}
\tilde{p}_0=\frac{1}{\lambda}\left(\delta(x)+ \frac{d}{dx}(U'+D\frac{d}{dx})\tilde{p}_0\right)
\end{equation}
in \e{j1_2} and integrating a few times by part, the following expression 
for the first integral in \e{j1_2} is obtained
\begin{eqnarray}
\int_x^{x_E} dy e^{U(y)/D} \tilde{p}_0(y,\lambda)&=&\frac{1}{\lambda}\Bigg(\Theta(-x)+D \Big[ e^{U_E/D} \tilde{p}_{0,E}'- e^{U(x)/D} \tilde{p}_0'(x)\Big] \nonumber\\
&& \;\;\;\;\;\;\;\;\;\; +\int_x^{x_E} dy\;\; U''(y) e^{U(y)/D} \tilde{p}_0(y) \Bigg) 
\end{eqnarray}
where an index ``E'' means that the respective function is taken at
$x=x_E$.  For the specific potential \e{pot}, this yields (using \e{rho_def})
\begin{equation}
\int_x^{x_E} dy e^{U(y)/D} \tilde{p}_0(y,\lambda)=\frac{1}{\lambda-b}\Bigg(\Theta(-x)- e^{U_{E}/D} \rho_0(\lambda) - D e^{U(x)/D} \tilde{p}_0'(x) \Bigg). 
\end{equation}
Here, $\rho_0(\lambda)$ denotes the Laplace transform of the FPT
density for the unperturbed system. Inserting this formula into \e{j1}
the following quadrature formula is obtained
\begin{equation}
\label{j1_inti}
j_1(x_E)=\frac{\lambda/D}{\lambda-b}\Bigg(e^{U_E/D}\rho_0(\lambda)
\int_{-\infty}^{x_E} dx e^{-U(x)/D} - \int_{-\infty}^{0} dx
e^{-U(x)/D} \Bigg).
\end{equation}
One may repeat the whole derivation for arbitrary $x$ (this is needed
in the calculation of $j_2(x_E)$), yielding
\begin{eqnarray}
j_1(x)&=&\bigg(\frac{1}{\lambda-b}\hat{\cal T}_0-\hat{\cal T}_1\bigg) P_0(x,t)+\nonumber \\
&& \hspace{-5em}+\frac{\lambda/D}{\lambda-b}\Bigg(e^{U_E/D}\rho_0(\lambda) \int_{-\infty}^{x} dy e^{-U(y)/D} - \int_{-\infty}^{x} dy e^{-U(y)/D}\Theta(-y) \Bigg).
\end{eqnarray}
Insertion into \e{j2} and a few manipulations of the occurring multiple 
integrals leads to the following correction of the second moment
\begin{eqnarray}
\label{j2_inti}
j_2(x_E)&=&-2\bigg(\frac{\lambda}{\lambda-b}+\lambda\frac{d}{d\lambda}\bigg) \frac{j_1(x_E)}{\lambda}+ \nonumber\\ 
&&  + \frac{2/D^2}{\lambda-b}\Bigg(\rho_0(\lambda) e^{U_E/D} \int_{-\infty}^{x_E} dx\;\; e^{-U(x)/D} I^2(x) -\nonumber\\
&&- \int_{-\infty}^{0} dx\;\; e^{-U(x)/D} I^2(x)-e^{U(0)/D} I(0)  \int_{0}^{x_E} dx\;\; I(x)\Bigg)
\end{eqnarray}
where 
\begin{equation}
I(x):=e^{U(x)/D} \int_{-\infty}^{x} dy\;\; e^{-U(y)/D}. 
\end{equation}
\subsection{Formulas for the linear potential case} 
I now turn to the specific case of a linear potential, which is
particularly simple. Assuming $b=0$ and $a>0$, I have for $\varepsilon=0$
(see, e.g. \cite{Hol76})
\begin{eqnarray}
\label{mean0_pif}
\lr{T}_0&=& \frac{x_E}{a},\\
\label{var0_pif}
\lr{\Delta T^2}_0&=&2 \frac{D x_E}{a^3},\\
\label{cv0_pif}
CV_0&=& \sqrt{\frac{2 D}{x_E a}},
\end{eqnarray}
and 
\begin{eqnarray}
\label{charfunc_pif}
\rho_0(\lambda)&=&\exp\left[\frac{x_E(a-\sqrt{a^2+4\lambda D})}{2D}\right]\\
\label{I_lin}
I(x)&\equiv& D/a.
\end{eqnarray}
Inserting the latter expressions into eqs.~(\ref{j1_inti}) and (\ref{j2_inti})
yields the following linear corrections of the first and second
moment, respectively
\begin{eqnarray}
\label{j1_pif}
j_1(x_E)&\!=\!&\frac{1}{a}\left(e^{\textstyle\frac{x_E(a-\sqrt{a^2+4\lambda D})}{2D}}-1\right),\\
\label{j2_pif}
j_2(x_E)&\!=\!&\frac{2}{a^2}\left\{\left(\frac{x_Ea}{\sqrt{a^2+4\lambda D}}+\frac{D}{a}\right) e^{\textstyle\frac{x_E(a-\sqrt{a^2+4\lambda D})}{2D}}-x_E-\frac{D}{a}\right\}. \nonumber\\
\end{eqnarray}
Hence, the mean and the variance of the first passage time in a linear
potential under the influence of a weak exponential driving are given by the
fairly simple expressions
\begin{eqnarray}
\label{mean_pif}
\lr{T}&\!=\!& \frac{x_E}{a}-\frac{\varepsilon}{a}\left(1- e^{\textstyle\frac{x_E(a-\sqrt{a^2+4\lambda D})}{2D}}\right),\\
\label{var_pif}
\displaystyle\lr{\Delta T^2}&\!=\!& 2D \frac{x_E-\varepsilon}{a^3}+\frac{2\varepsilon}{a^2}\left(\frac{x_E a}{\sqrt{a^2+4\lambda D}}+\frac{D}{a}-x_E \right) e^{\textstyle\frac{x_E(a-\sqrt{a^2+4\lambda D})}{2D}}.  \nonumber\\
\end{eqnarray}
It can be easily seen that the corrections to the moments
\e{j1_pif} and \e{j2_pif} are negative if $\varepsilon>0$ 
(I recall that $a,D,$ and $\lambda$ are positive).  This makes sense,
since a positive force toward threshold will always diminish the first
passage time and hence also its moments. Furthermore, the correction
of the variance is also negative for $\varepsilon$ being positive.

For very small decay rate ($\lambda<<1/\lr{T}_0$), an expansion
of \e{mean_pif} and \e{var_pif} in $\lambda$  yields
\begin{eqnarray}
\label{mean_pif_small_lam}
\lr{T} &\to & \lr{T}_0-\frac{x_E}{a^2}\lambda\varepsilon ,\;\;\;\;\; \lambda<<1/\lr{T}_0 \\
\label{var_pif_small_lam}
\lr{\Delta T^2} &\to & \lr{\Delta T^2}_0-6\frac{x_E D \varepsilon\lambda}{a^4},\;\;\;\;\; \lambda<<1/\lr{T}_0 
\end{eqnarray}  
This is also obtained if in the formulas of the unperturbed system
\e{mean0_pif} and \e{var0_pif} the bias $a$ is replaced by
$a+\varepsilon\lambda$ and the formulas are expanded up to linear order in
$\varepsilon$. Hence the static limit confirms the result of the perturbation
calculation.\\
Considering the limit of large decay rate ($\lambda\to \infty$), I
note that the exponential function in eqs.~(\ref{mean_pif}) and
~(\ref{var_pif}) can be neglected and I obtain the simple limit
\begin{eqnarray}
\label{mean_pif_lalam}
\lr{T}\displaystyle &\to& \frac{x_E-\varepsilon}{a} ,\;\;\;\;\;\lambda\to \infty\\
\label{var_pif_lalam}
\lr{\Delta T^2}&\displaystyle\to& 2D \frac{x_E-\varepsilon}{a^3},\;\;\;\;\;\lambda\to \infty
\end{eqnarray}
In the limit $\lambda\to \infty$ the driving acts as a $\delta$ spike
at $t=0$ with amplitude $\varepsilon$, which leads to a modified initial
point $x(t=0)=\varepsilon$. To check this, one can again use the formulas for
the unperturbed system by replacing the distance between initial point
and absorbing boundary (which was $x_E$) by $x_E-\varepsilon$. This leads
{\em exactly} to \e{mean_pif_lalam}, \e{var_pif_lalam},
i.e. in the limit $\lambda\to\infty$ the result of the perturbation
calculation is exact.\\
\subsection{Formulas for the parabolic potential case}
Mean, variance, and Laplace transform of the FPT density for the case
of the parabolic potential and $\varepsilon=0$ can be written as follows
(see, for instance, Refs.~\cite{Sie51,Hol76,LinLSG02}
\begin{eqnarray}
\label{mean0_lif}
\langle T \rangle_0&=&\frac{\sqrt{\pi}}{b}\int\limits_{x_-}^{x_+} dy\; e^{y^2} \mbox{erfc}(y), \\ 
\label{var0_lif}
\langle \Delta T^2 \rangle_0&=& \frac{2\pi}{b^2} \int\limits_{x_-}^\infty dy\; e^{y^2} [\mbox{erfc}(y)]^2 \int\limits_{x_-}^y dz\; e^{z^2} \Theta\left(x_+-z\right)  
\end{eqnarray}
and
\begin{equation}
\rho_0(\lambda)=e^{-\delta /2 }\frac{\pcf{-\lambda/b}{x_+\sqrt{2}}}{\pcf{-\lambda/b}{ x_-\sqrt{2}}}
\end{equation}
where 
\begin{equation}
x_-=\frac{a-b x_E }{\sqrt{2 D b}},\;\;\;\;
x_+=\frac{a}{\sqrt{2 D b}},\;\;\;\;
\delta=x_-^2-x_+^2.
\end{equation}
In these expressions, erfc$(x)$ denotes the complementary error
function and $\pcf{\alpha}{z}$ is the parabolic cylinder function
\cite{AbrSte70}. The auxiliary function $I(x)$ reads now
\begin{equation}I(x)=\sqrt{\frac{\pi D}{2 b}} \exp\left[\frac{(a-xb)^2}{2Db}\right] \mbox{erfc}\!\left(\frac{a-xb}{\sqrt{2Db}}\right)
\end{equation} 
Using this function, I obtain the following correction to the mean first
passage time out of a  parabolic potential
\begin{equation}
\label{j1_lif}
j_1(x_E)=\sqrt{\frac{\pi}{2 b D}}\frac{\lambda}{\lambda-b}e^{x_+^2} \left[e^\delta \rho_0(\lambda) \mbox{erfc}\!\left(x_-\right)-  \mbox{erfc}\!\left(x_+\right) \right]
\end{equation}
By means of the derivative of $j_1(x_E)$ with respect to $\lambda$, the correction to
the second moment can be brought in the following form
\begin{eqnarray}
\label{j2_lif}
j_2(x_E)&=&-2\lambda\left(\frac{d}{d\lambda}+\frac{1}{\lambda-b}\right) \frac{j_1(x_E)}{\lambda} - \sqrt{\frac{2}{D b^3}} \frac{\lambda \pi}{\lambda-b} e^{x_+^2}\times \nonumber\\
&& \times\left[\;\;\int\limits_{x_+}^\infty dx\; e^{x^2} \mbox{erfc}^2\!\left(x\right) - e^\delta \rho_0(\lambda) \int\limits_{x_-}^\infty dx\; e^{x^2} \mbox{erfc}^2\!\left(x \right)\nonumber+ \right. \\
&& \left.\;\;\;\;\;+\mbox{erfc}\!\left(x_+\right) \int\limits_{x_-}^{x_+} dx\; e^{x^2} \mbox{erfc}\!\left(x\right) \right]
\end{eqnarray}
The resulting formulas for the FPT's mean and variance are given by 
\begin{eqnarray}
\label{mean_lif}
\lr{T}&=&\lr{T}_0+\frac{\varepsilon\lambda}{\lambda-b}\sqrt{\frac{\pi}{2 b D}}e^{x_+^2} \left[e^\delta \rho_0(\lambda) \mbox{erfc}\!\left(x_-\right)- \mbox{erfc}\!\left(x_+\right) \right]\\[1em]
\label{var_lif}
\lr{\Delta T^2}&=& \lr{\Delta T^2}_0-\frac{\varepsilon\lambda}{\lambda-b} \sqrt{\frac{2\pi}{D b}} e^{x_+^2} \Bigg[ e^\delta\mbox{erfc}\!\left(x_-\right)\big[\rho_0'(\lambda)+ \lr{T}_0 \rho_0(\lambda)\big]  \;\;\; +\nonumber \\
&& +\frac{\sqrt{\pi}}{b} \int\limits_{x_-}^\infty dx\; e^{x^2} \mbox{erfc}^2(x) \big[\Theta(x-x_+)-e^\delta \rho_0(\lambda)\big] \Bigg] 
\end{eqnarray}
where $\rho_0'(\lambda)$ denotes the derivative \footnote{Since there
is no simple analytical expression for this derivative, I perform it
numerically:
$\rho_0'(\lambda)=[\rho_0(\lambda+\varepsilon_\lambda)-\rho_0(\lambda)]/\varepsilon_\lambda$
with $\varepsilon_\lambda=10^{-5}$.} of $\rho_0(\lambda)$ with respect to $\lambda$.\\
Again, the limits of small and large-$\lambda$ may be considered. For
$\lambda\to 0$, it is easily seen from \e{moments_rho} that $\rho_0\to
1$ and $\rho_0'\to -\lr{T}_0$. With this I obtain
\begin{eqnarray}
\lr{T}&\approx&\lr{T}_0-\frac{\varepsilon\lambda}{b}\sqrt{\frac{\pi}{2 b D}} \left[e^{x_-^2} \mbox{erfc}\!\left(x_-\right)- e^{x_+^2} \mbox{erfc}\!\left(x_+\right) \right],\;\;\; \lambda<<1/\lr{T}_0\nonumber\\
\label{mean_lif_small_lam}\\
\lr{\Delta T^2}&\approx& \lr{\Delta T^2}_0-\frac{\varepsilon\lambda\pi}{b^2}\sqrt{\frac{2}{ D b}} \times\nonumber\\
&&\hspace*{-2em}\times \Bigg(\;\int\limits_{x_-}^\infty dx\; e^{x^2+x_-^2} \mbox{erfc}^2(x)-\int\limits_{x_+}^\infty dx\; e^{x^2+x_+^2} \mbox{erfc}^2(x)\Bigg),\;\;\; \lambda<<1/\lr{T}_0 \nonumber\\
\label{var_lif_small_lam}
\end{eqnarray}
The terms proportional to $\varepsilon\lambda$ are also obtained by taking
the derivative of mean or variance of the unperturbed system given in
eqs.~(\ref{mean0_lif}) and (\ref{var0_lif}) with respect to parameter $a$. For very
slow driving, the perturbation acts as a static change in the bias
parameter. Consequently, the perturbation calculation leads to the
same result like a linearization of mean and variance with respect to a small
change in the bias parameter.\\
In case of infinite $\lambda$, the characteristic function $\rho_0$ and its
derivative $\rho_0'$ approach zero yielding the following simplified 
expressions for mean and variance
\begin{eqnarray}
\label{mean_lif_lalam}
\lr{T}\to \lr{T}_0-\varepsilon\sqrt{\frac{\pi}{2 b D}} e^{x_+^2} \mbox{erfc}\!\left(x_+\right),\;\;\;\;\;\lambda\to \infty\\[1em]
\label{var_lif_lalam}
\lr{\Delta T^2}\to \lr{\Delta T^2}_0-\varepsilon\sqrt{\frac{2}{ D b^3}}\pi e^{x_+^2} \int\limits_{x_+}^\infty dx\; e^{x^2} \mbox{erfc}^2(x) ,\;\;\;\;\;\lambda\to \infty
\end{eqnarray}
Again, what physically happens in this case is a shift of the initial
point from $x(t=0)=0$ to $x(t=0)=\varepsilon$ since the driving $\varepsilon \lambda
e^{-\lambda t}$ acts as a $\delta$ spike at the initial
time. Consequently, the above results are also obtained if in
eqs.~(\ref{mean0_lif}) and (\ref{var0_lif}) $x_+$ (the only term where
the initial point enters) is replaced by $(a-b
\varepsilon)/\sqrt{2D b}$ and the expressions are expanded to linear order in $\varepsilon$.
This in turn, is another check that the results achieved cannot be completely
wrong. 
\section{Mean and variance of the FPT: Comparison to simulations}
\label{sec:simus}
As a verification of the specific results derived in the previous
section, I consider the mean, the variance and the coefficient of
variation (CV) of the FPT in a linear and in a parabolic potential. It
will become apparent that an exponential driving of these systems can
result in a remarkable behavior of the first two cumulants.\\
For all data shown, I use a weak positive driving amplitude of
$\varepsilon=0.05$, an intermediate noise intensity $D=0.1$, and the
absorbing boundary to be at $x_E=1$. Furthermore, two different sets
of potential parameters are inspected: (i) $a=1, b=0$ for the linear
potential and (ii) $a=0.8, b=1$ in case of a parabolic
potential. The latter choice implies a significantly different FPT
statistics since here the escape from the potential minimum at $x=0.8$
will dominate the passage time (for a large value of $a$, the
potential minimum is beyond the absorbing boundary and the FPT
statistics will be akin to the linear-potential case). I compare the
analytical results derived above to simulation results, that were
obtained with a simple Euler procedure in case of a linear potential
(time step was $\Delta t=10^{-4}$ and $10^6$ passage times were
simulated), and a modified Euler procedure \cite{Hon89} for the
parabolic potential (time step was $\Delta t=10^{-3}$ and $10^6$
passage times were simulated).  In all curves data for $\varepsilon=0$ for
which I know the exact values of all quantities are shown for the sake
of illustration and also to demonstrate the validity and accuracy of
the numerical simulation procedure.\\
Since the agreement between theory and simulations is excellent, I do
not have to discuss it at length; I can regard it as a satisfying
first confirmation of the presented analytical approach. Concerning
the agreement between theory and simulations at other parameter sets
(e.g. larger driving amplitude and smaller or larger noise intensity),
I restrict myself to the following brief statement: the perturbation
result for the correction to a moment yields satisfying quantitative
agreement with the simulations as long as the correction is small
compared to the respective unperturbed moment. In general the theory
will work best for intermediate up to large noise intensities since
with a non-weak noise the effect of an additional weak driving will be
only moderate and the first (linear) correction term will suffice.
Note, however, that for systems without a potential barrier between
initial point and absorbing boundary (like, for instance, the linear
potential), the theory works at  weak noise, too.\\
In the remainder of this section I focus on the statistical features
of the exponentially driven first passage process as they are
reflected by mean, variance, and CV as functions of the decay rate
$\lambda$.\\
\subsection{Biased random walk with exponential forcing}
It can be expected that a positive exponentially decaying forcing
leads to a decrease of the mean FPT. For the linear potential (i.e. a
biased random walk) this decrease is a monotonous function of $\lambda$
as shown in \bi{mean_pif}. At small $\lambda$ the correction is
proportional to $\lambda$ according to the static approximation
\e{mean_pif_small_lam} (shown by the dot-dashed line in
\bi{mean_pif}). In this range of $\lambda$ the driving is effectively
static with amplitude $\lambda\varepsilon$ meaning that its decay
occurs on a time scale that is far beyond the mean FPT. In other
words, a part of the driving's power is ``wasted'' because $s(t)$
still drives the system long after most realizations have been
absorbed at $x=x_E$. For intermediate values of $\lambda$, the decay
of the driving force takes place much earlier and thus, its effect on
the mean is less than that of a static driving. Finally, in the
large-$\lambda$ limit the mean saturates according to
\e{mean_pif_lalam} at the value corresponding to a change of the
initial point in the unperturbed system. I note that the monotonous
behavior of $\lr{T}$ as a function of $\lambda$ differs from what was
found for {\em periodically} driven linear systems in
Refs.~\cite{FleHav88,GitWei95}. In the latter case, minima
\cite{FleHav88} or maxima \cite{GitWei95} of the mean FPT vs the
driving frequency were observed for different boundary and initial
conditions.

\begin{figure}[h!]
\centerline{\parbox{9cm}{\epsfig{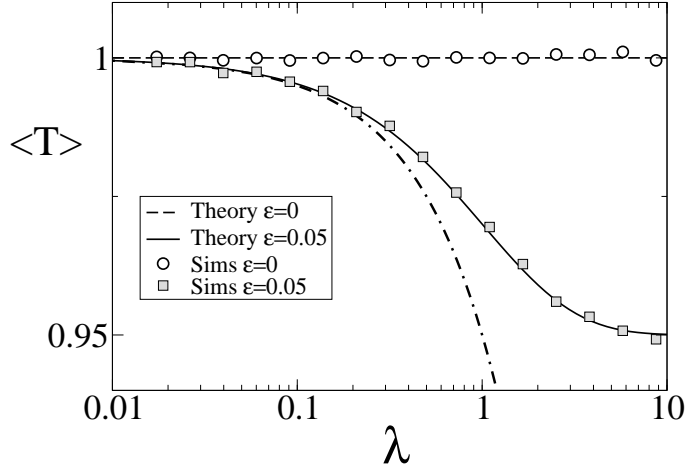}}}
\caption{\small\label{fig:mean_pif}
The mean of the first passage time vs decay rate of the exponential
driving in case of a linear potential.  Simulations (symbols) and
theory (lines) \e{mean_pif} with the indicated values of the driving amplitude.   The dot-dashed line illustrates the
slow-driving approximation \e{mean_pif_small_lam}.}
\end{figure}

The variance of the FPT (\bi{var_pif}) is always below that of the
unperturbed case. It shows, remarkably, a nonmonotonous behavior as a
function of the decay rate. For the parameter set used in
\bi{var_pif}, the variance attains a minimum at $\lambda\approx 1.6$.
It is possible to calculate the exact location of this minimum from
\e{var_pif} and express it solely by means of the mean $\lr{T}_0$ and 
the squared coefficient of variation (denoted for brevity by
$R=CV_0^2$) of the unperturbed system
\begin{equation} 
\lambda_{min}=\frac{\sqrt{1+4 R-2 R^2}-3 R^2/2+4 R-1}{\lr{T}_0 R (2-R)^2}
\end{equation}

\begin{figure}[h!]
\centerline{\parbox{9cm}{\epsfig{file=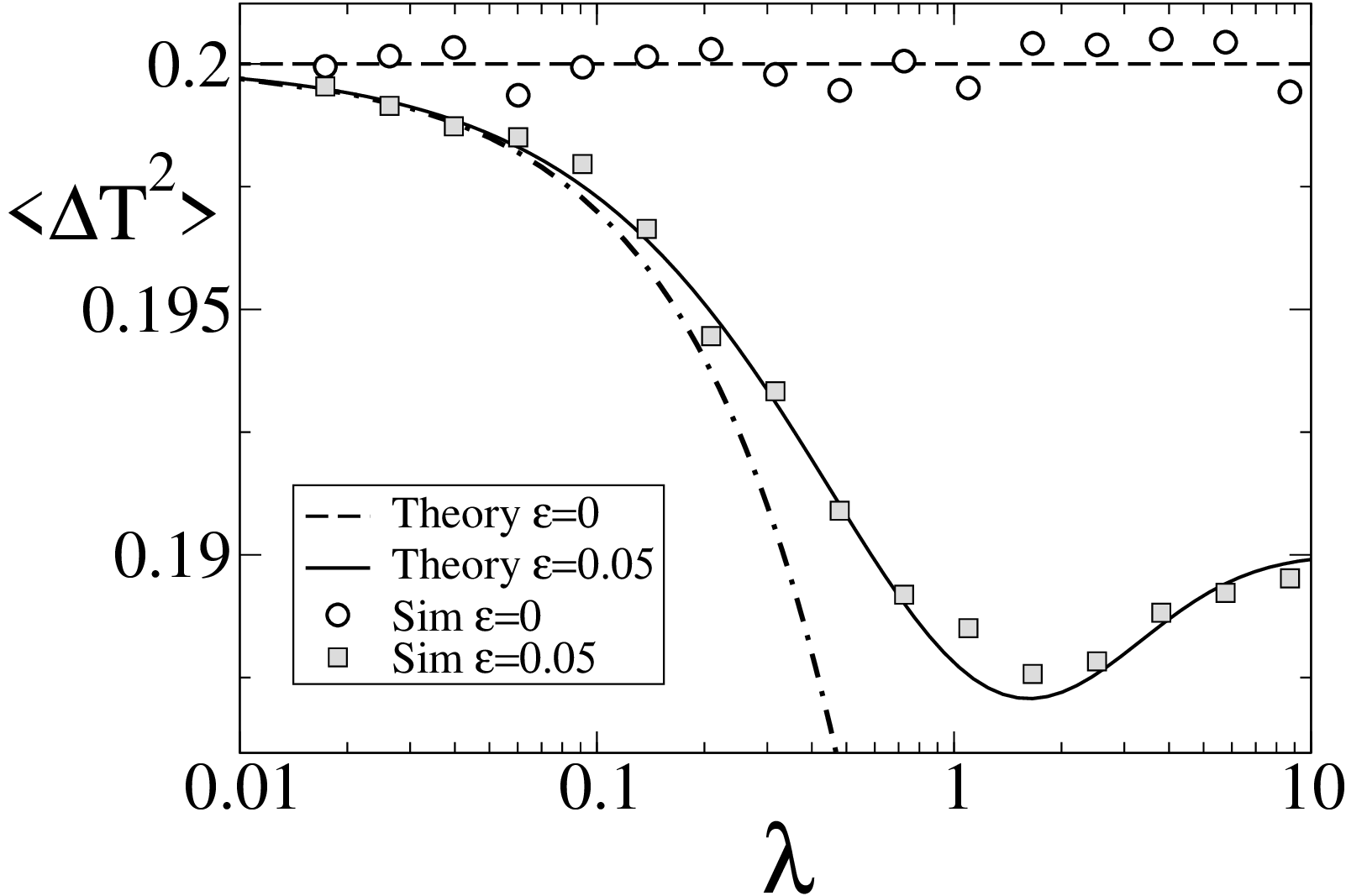,width=9cm,angle=0}}}
\caption{\small\label{fig:var_pif}
The variance of the first passage time vs decay rate of the
exponential driving in case of a linear potential.  Simulations
(symbols) and theory \e{var_pif} (lines) with the indicated values of the driving
amplitude.   The dot-dashed line illustrates the
slow-driving approximation \e{var_pif_small_lam}.}
\end{figure}

A minimum in the variance does not occur for an arbitrary parameter
set but if and only if
\begin{equation}
\label{cond_min}
R<2 \;\; \Rightarrow \;\; CV_0=\sqrt{2D/(a x_E)} <\sqrt{2},
\end{equation}
i.e. for sufficiently weak noise intensity or large bias $a$. If the
condition \e{cond_min} is met, the value at which the minimum is
attained is an increasing function of $R$ diverging at $R=2$ and
saturating for small values of the CV (i.e. $D\to 0$ or $a\to \infty$)
at
\begin{equation}
\label{lam_min_smallD}
\lambda_{min} \to \frac{3}{2} \frac{1}{\lr{T}_0}\;\;\; \mbox{as} \;\; CV_0\to 0.
\end{equation}
The occurrence of the minimum seems to be related to the fact that a
time-dependent bias reduces the variability more strongly than a shift
in the initial point (corresponding to the limit $\lambda\to\infty$)
does. This gives raise to the drop of the variance as $\lambda$ is
decreased starting in the large-$\lambda$ limit. The amplitude of the
time-dependent driving, however, depends on $\lambda$, too, so its
effect on the dynamics gets weaker by further decreasing $\lambda$ and
the variance starts to increase again. Accordingly, using the
exponential driving {\em without} the prefactor $\lambda$ (i.e.,
without normalizing the driving's intensity) yields a variance that
grows monotonously with $\lambda$ (not shown). Therefore, the minimum
of the variance is merely based on two competing effects, namely, the
greater impact of a slow driving (compared to a fast one) on the
reduction of the variance and the dependence of the variance on the
driving's amplitude (i.e. $\varepsilon\lambda$).\\
\begin{figure}[h!]
\centerline{\parbox{9cm}{\epsfig{file=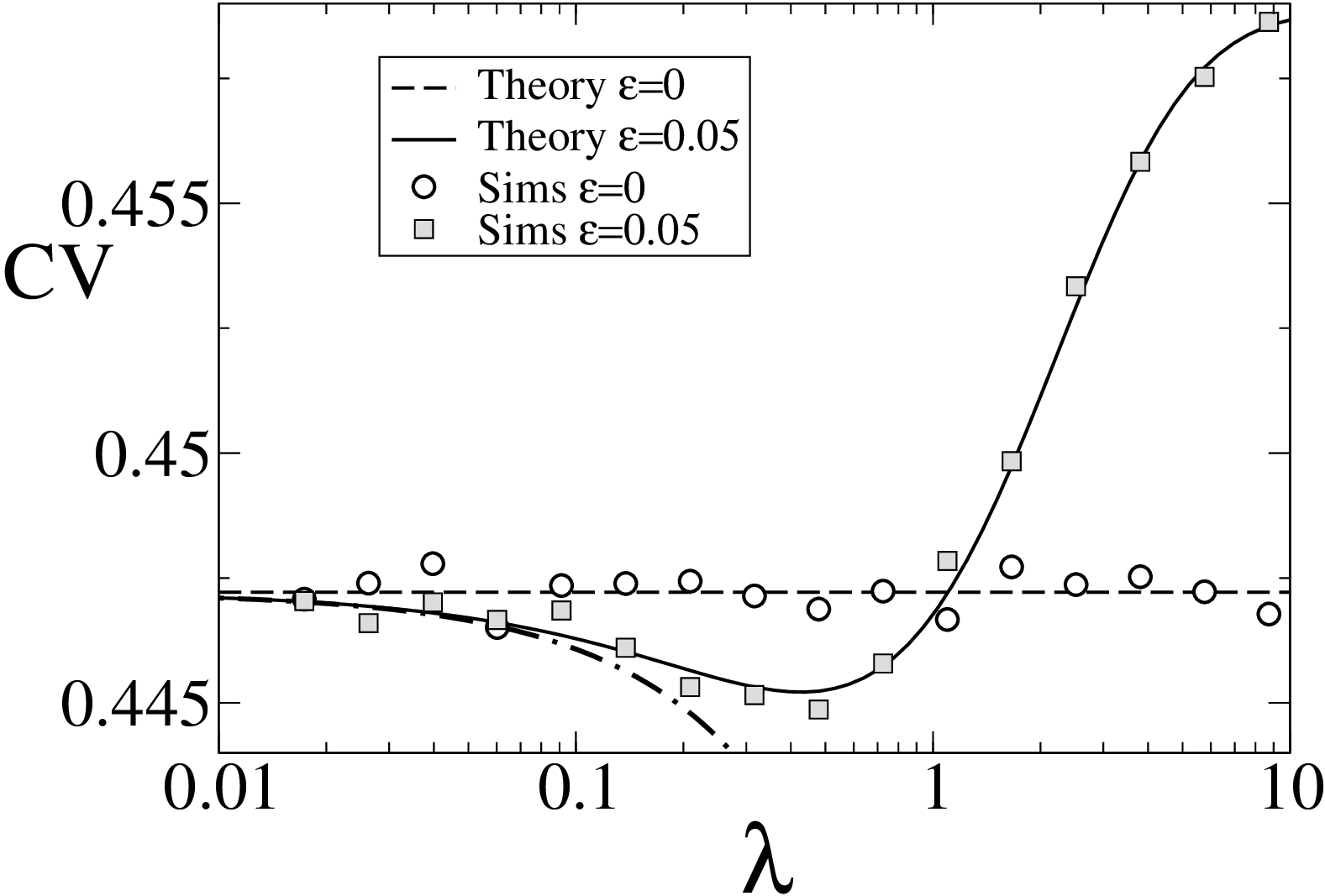,width=9cm,angle=0}}}
\caption{\small\label{fig:cv_pif}
The coefficient of variation of the first passage time vs decay rate
of the exponential driving in case of a linear potential. Simulations
(symbols) and theory using \e{cv_def}, \e{mean_pif}, and \e{var_pif}
(lines) with the indicated values of the driving amplitude. The
dot-dashed line illustrates the slow-driving approximation using
\e{cv_def}, \e{mean_pif_small_lam}, and \e{var_pif_small_lam}.}
\end{figure}

The minimum in the variance vs $\lambda$ could be interpreted as an
``optimal'' decrease in variability due to an exponential
driving. Things look different, though, from the view point of {\em
relative} variability as it is quantified by the CV (cf.
\bi{cv_pif}). First of all, depending on the value of $\lambda$, the
CV can be both larger or smaller than in the unperturbed case. For the
value of $\lambda$ at which the variance attains a minimum the CV is
{\em larger} than in the unperturbed case and is thus far from being
``optimal''.  The fact that the effect of the exponential driving on
the CV can be both positive or negative can be understood by looking
at the CV in the limit cases of small and large decay rate where it
corresponds to the CV of the unperturbed system with rescaled
parameters.  The latter depends on the inverse of the product $a
x_E$. A static increase of the bias (replacing $a$ by $a+\varepsilon\lambda$
which is the effect of a slow driving) will thus lead to a decrease in
CV compared to the unperturbed case (cf. the static approximation in
\bi{cv_pif}).  In contrast, diminishing the difference between initial
point and absorbing boundary (replacing $x_E$ by $x_E-\varepsilon$ which is
the effect of the forcing for $\lambda\to\infty$) leads to a higher CV
than in the unperturbed case. Interpolating between the two limit
cases will inevitably lead to at least one minimum of the CV vs
$\lambda$. For the parameter set used in \bi{cv_pif}, this minimum is
attained at a decay rate that is smaller than the inverse mean FPT of
the unperturbed system.\\ 
In conclusion, already in the simple linear potential case, the effect
of an exponential driving can be fairly involved. For the behavior of
mean, variance, and CV as functions of the decay rate, it was essential
that I have used  a constant-intensity scaling of the driving function.
\subsection{Escape out of a parabolic potential with exponential forcing}
With $b> 0$ there is a true state dependence on the right hand side of
the dynamics \e{langevin}.  The state variable is attracted toward the
potential minimum at $x=a/b$.  If this rest state is far beyond the
absorbing boundary $x_E$ (i.e. $x_E\ll a/b$), the FPT statistics will
be similar to that of the biased random walk. In the following I
choose, however, $b=1$, $a=0.8$, and $x_E=1$ such that $a/b<x_E$.
With this choice the FPT problem is significantly different from the
linear potential previously discussed, since in order to reach the
absorbing boundary at $x_E=1$ the state variable $x(t)$ has to be
driven by a sufficient noise to accomplish the escape out of the
potential minimum.\\
\begin{figure}[h!]
\centerline{\parbox{9cm}{\epsfig{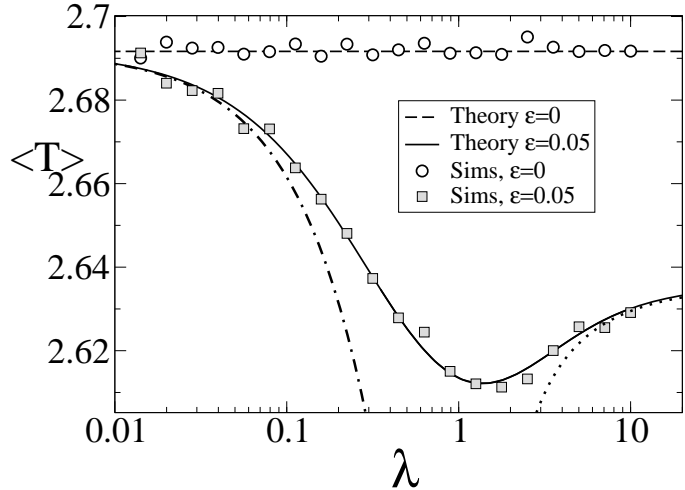}}}
\caption{\small\label{fig:mean_lif}
The mean of the first passage time vs decay rate of the exponential
driving in case of a parabolic potential.  Simulations (symbols) and
theory \e{mean_lif} (lines) with the indicated values of the driving
amplitude.  The dot-dashed line illustrates the slow-driving
approximation \e{mean_lif_small_lam}; the dotted line is the extended
large-$\lambda$ approximation \e{mean_lif_lalam2}.  }
\end{figure}
In case of a parabolic potential, already the mean FPT depends
nonmonotonously on the driving's decay rate $\lambda$ (\bi{mean_lif});
it attains a minimum for $\lambda\approx 1.5$ which stands in marked
contrast to the linear case. The reason for the occurrence of this
minimum is the state dependence of the dynamics \e{langevin} as I will
show now. First of all, starting at $\lambda=0$, the mean FPT
decreases linearly with $\lambda$ in accord with the static
approximation shown by the dot-dashed line in \bi{mean_lif}. This is
completely equivalent to the linear potential case. Second, the mean
FPT also drops if the decay rate is decreased starting from the
large-$\lambda$ limit. In other words, the limiting value is
approached from below. The behavior of the mean in these two limits
implies the occurrence of at least one minimum.\\
In order to understand why the large-$\lambda$ limit is approached from below,
I consider a large but finite value of $\lambda$ and a time $t$ for which
\begin{equation}
\label{condi_appro}
\frac{1}{\lambda}\ll t \ll \lr{T}_0.
\end{equation}
For such times it can be taken for granted that the driving has
practically decayed to zero but on the other hand it is highly
unlikely that a realization has already reached $x_E$. In this case
the effect of the driving can be inferred from the free solution of
\e{langevin} for a parabolic potential with initial value $x(t=0)=0$
which can be written as follows
\begin{equation}
x(t)=\varepsilon\left(1+\frac{b}{\lambda-b}\right)e^{-b t}-\varepsilon\frac{\lambda}{\lambda-b}e^{-\lambda t} + \int\limits_0^t d\tilde{t} e^{b(\tilde{t}-t)} [a+\sqrt{2D}\xi(\tilde{t})]
\end{equation}
If the condition \e{condi_appro} is met, the second term can be
neglected and the solution reads
\begin{equation}
\label{langevin_appro}
x(t)=\varepsilon\left(1+\frac{b}{\lambda-b}\right)e^{-b t}+ \int\limits_0^t
d\tilde{t} e^{b(\tilde{t}-t)} [a+\sqrt{2D}\xi(t)]
\end{equation}
This approximate solution is, however, equivalent to the unperturbed
dynamics with an initial point at $x(t=0)=\varepsilon[1+b/(\lambda-b)]$. The
equivalence holds true for a time $t$ obeying \e{condi_appro} {\em and
any time larger than this time}.  In other words, for $t\gg 1/\lambda$
the realization of the original process and that of the unperturbed
process with the modified initial condition differ only by a small
exponential contribution. Consequently, also the FPT statistics of
both processes will be the same provided that a successful escape
toward $x_E$ is highly unlikely for short times at which
\e{langevin_appro} does not hold true.\\
For $\lambda\to\infty$, the initial condition approaches the value
$x(t=0)\to\varepsilon$ (see the discussion around \e{mean_lif_lalam}). For a
large but finite value of $\lambda$, the shift in the initial point
will be {\em larger} than in the latter limit and thus the mean will
be more strongly decreased than in the strict limit
$\lambda\to\infty$.  To obtain an explicit formula showing this drop,
the mean of the unperturbed system with modified initial point is
linearized with respect to $\varepsilon$ (this is not strictly necessary but
consistent with the linear approach used throughout this paper)
yielding
\begin{equation}
\label{mean_lif_lalam2}
\lr{T}\approx\lr{T}_0-\varepsilon\left(1+\frac{b}{\lambda-b}\right) \sqrt{\frac{\pi}{2 b D}} e^{x_+^2} \mbox{erfc}\!\left(x_+\right),\;\;\;\;\;\lambda\gg1/\lr{T}_0
\end{equation}
Of course, this is also obtained by replacing the amplitude $\varepsilon$ in
\e{mean_lif_lalam} by the modified amplitude $\varepsilon[1+b/(\lambda-b)]$.  
The approximation is shown in \bi{mean_lif} by a dotted line; it
displays the drop of the mean with decreasing $\lambda$ at large decay
rate and agrees well with the full solution in this range.\\
I note that a nonmonotonous behavior of the mean was also found for a
system with parabolic potential and a {\em periodic} forcing in
ref.~\cite{BulEls96}. If the system is driven by $s(t)=\varepsilon
\cos(\omega t)$ the mean passes through maxima and - less
pronounced - minima when plotted as a function of the driving
frequency (cf. in particular Figs. 11,12, and 13 in
\cite{BulEls96}). There are two important differences between the exponential 
and periodic driving functions: (1) the periodic driving attains both
positive and negative values; (2) the amplitude of the periodic
forcing as considered in Ref.~\cite{BulEls96} is fixed and does not
depend on the time scale of the driving. The maxima and minima found
for periodic driving are true resonance phenomena.  In contrast, the
minimum in the mean FPT for exponential driving appears as a
compromise between the dependence of the driving's amplitude on the
decay rate (drop of the mean with increasing $\lambda$ at small
$\lambda$) and the stronger effect of a truly time dependent driving
on the state-dependent dynamics (drop of the mean with decreasing
$\lambda$ for $\lambda\gg1/\lr{T}_0$). I would like to point out that
the latter argument does not apply in case of a linear potential
(i.e., a state-independent force) because for $b=0$ the $\lambda$
dependent modification of the initial point in \e{langevin_appro}
vanishes. Hence, in this case we cannot infer the existence of a
minimum of $\langle T\rangle$ vs $\lambda$ and, in fact, it also does
not occur as was seen in the previous subsection.\\
\begin{figure}[h!]
\centerline{\parbox{9cm}{\epsfig{file=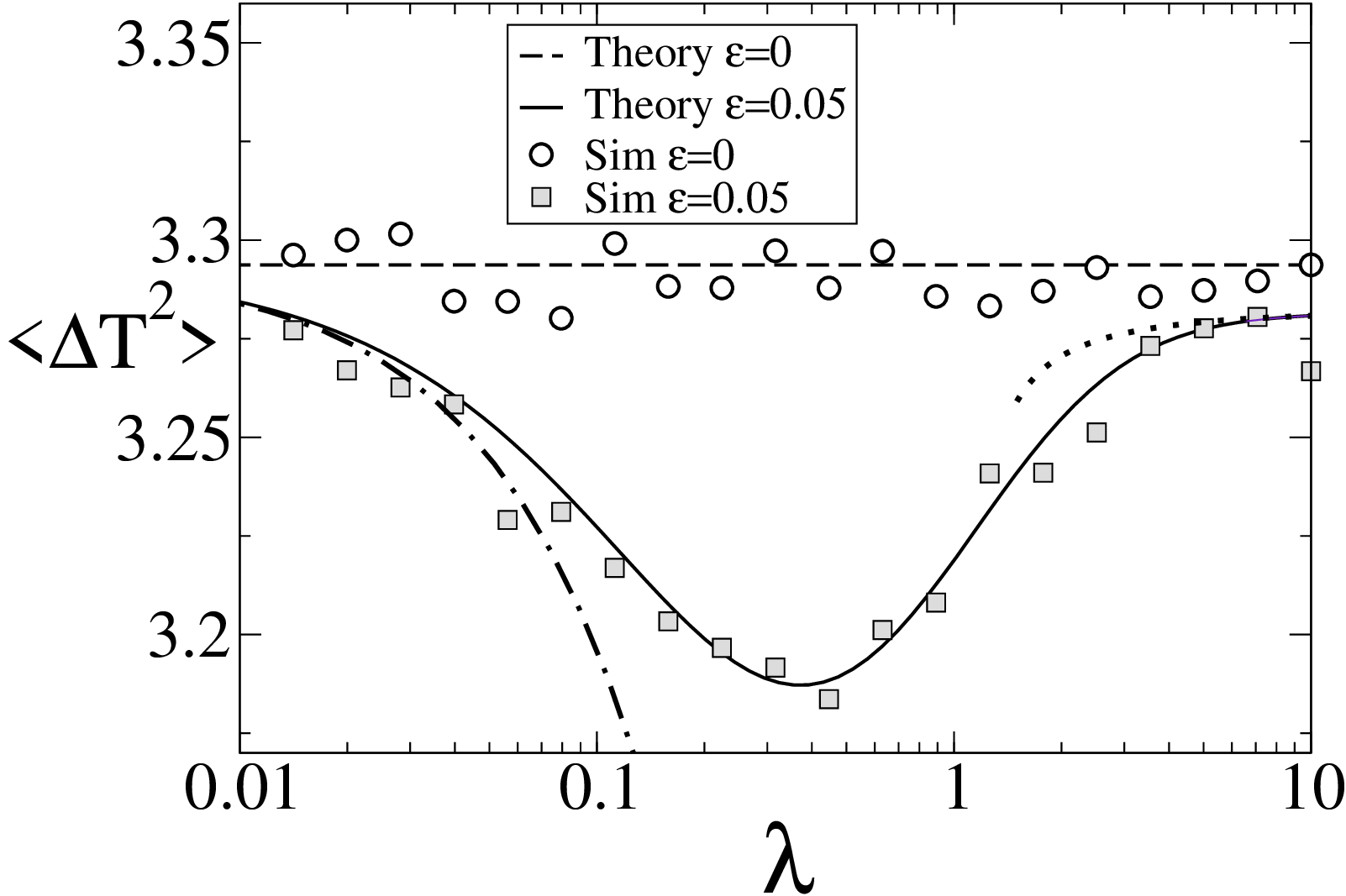,width=9cm,angle=0}}}
\caption{\small\label{fig:var_lif}
The variance of the first passage time vs decay rate of the
exponential driving in case of a parabolic potential.  Simulations
(symbols) and theory \e{var_lif} (lines) with the indicated values of
the driving amplitude.  The dot-dashed line illustrates the
slow-driving approximation \e{var_lif_small_lam}; the dotted line is
the extended large-$\lambda$ approximation \e{var_lif_lalam2}.}
\end{figure}

Turning to the variance depicted in \bi{var_lif}, I note that this
function also passes through a minimum vs $\lambda$ like in the linear
case. This minimum occurs at a smaller decay rate ($\lambda\approx
0.38$) than that of the mean FPT and remarkably close to the inverse
mean first passage time of the unperturbed system ($1/\lr{T}_0\approx
0.37$). Plotting the analytical solution \e{var_lif} for different
parameters revealed that this time-scale matching condition holds true
as long as the system is in the ``subthreshold'' regime, i.e. for
$a/b<x_E$ and weak up to moderate noise intensity. For larger noise
intensity and/or ``suprathreshold'' system parameters ($a/b>x_E$) the
minimum is attained at values larger than $1/\lr{T}_0$. For the
specific limit of weak noise and $a\to\infty$, one can expect the
location at the value found for the linear potential, namely
$\lambda_{min} \to 3/(2\lr{T}_0)$ which is indeed larger than the
inverse of the mean FPT in the unperturbed case.\\
The minimum of the variance can be understood by the same line of
reasoning as in case of the mean, i.e. by considering the behavior at
small and large-$\lambda$ which are determined by the static
approximation and by the effective solution \e{langevin_appro},
respectively. the latter leads - in complete analogy to the derivation
of \e{mean_lif_lalam2} - to the extended large-$\lambda$ approximation
of the variance
\begin{equation}
\label{var_lif_lalam2}
\lr{\Delta T^2}\!\approx\!\lr{\Delta T^2}_0-\varepsilon\left(1+\frac{b}{\lambda-b}\right)\sqrt{\frac{2}{D b^3}}\pi\!e^{x_+^2}\!\int\limits_{x_+}^\infty dx\; e^{x^2} \mbox{erfc}^2(x) ,\;\;\;\lambda\gg \frac{1}{\langle T\rangle_0}.
\end{equation}
This is shown in Fig.~\ref{var_lif} by the dotted line.  I note,
however, that the actual drop in variance at large $\lambda$ extends
over a much larger range where \e{var_lif_lalam2} does not hold true
anymore; the decrease of the variance in this range is also much
stronger than expected from \e{var_lif_lalam2}. The effect of a
temporally extended driving is thus much stronger than a change of the
initial point similarly to the case of a linear
potential. Furthermore, because the amplitude of the driving depends
on $\lambda$, the variance will drop for $\lambda\to 0$ to the value
of the unperturbed system. The occurrence of the minimum is therefore
mainly based on the different sensitivity of the FPT statistics with
respect to changes in the bias term $a$ and the initial point of the
passage and the $\lambda$ dependence of the driving amplitude due to
the constant-intensity scaling of the forcing.\\
\begin{figure}[h!]
\centerline{\parbox{9cm}{\epsfig{file=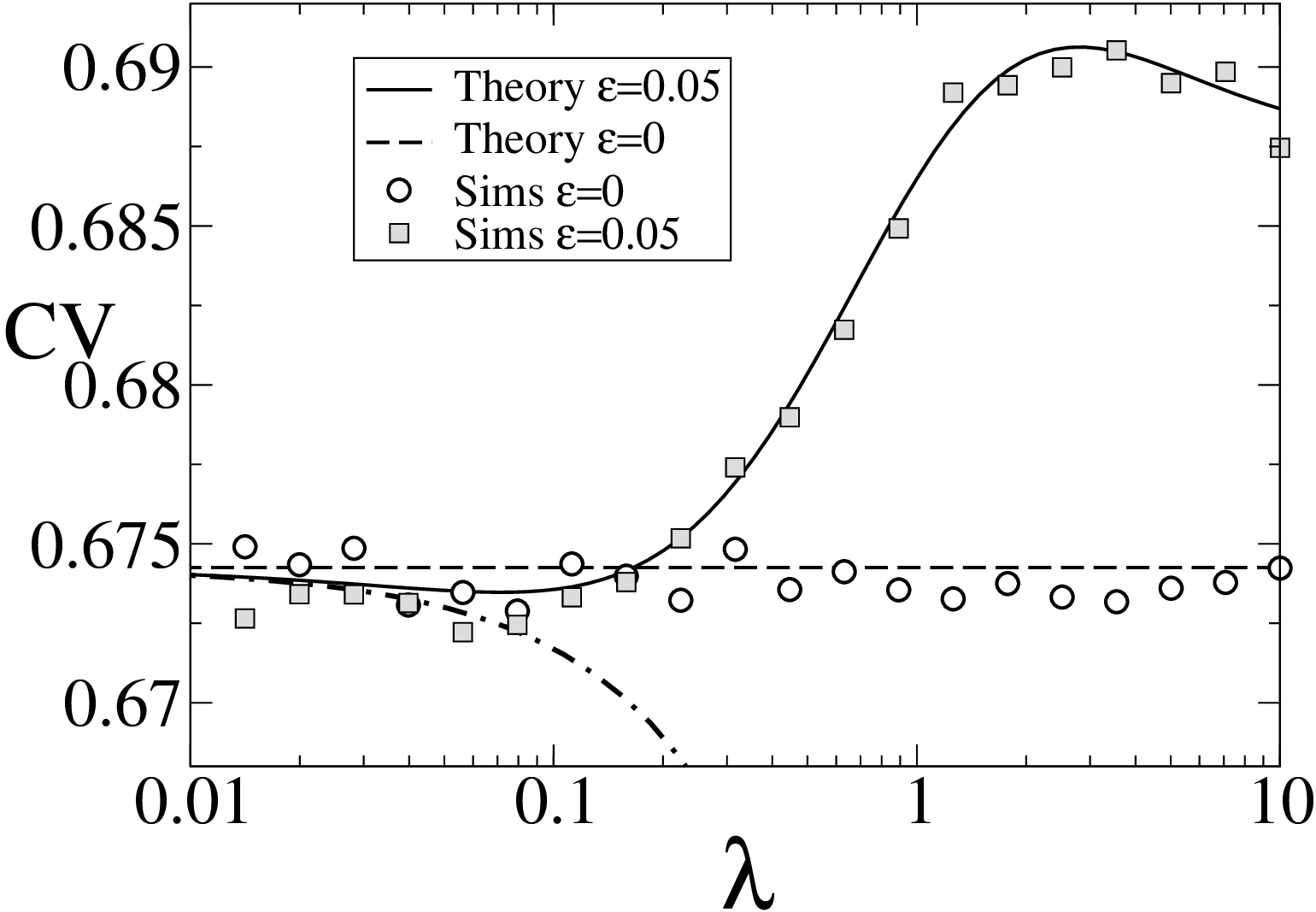,width=9cm,angle=0}}}
\caption{\small\label{fig:cv_lif}
The coefficient of variation of the first passage time vs decay rate
of the exponential driving in case of a parabolic potential.
Simulations (symbols) and theory using \e{cv_def}, \e{mean_lif}, and
\e{var_lif} (lines) with the indicated values of the driving
amplitude.  The dot-dashed line illustrates the slow-driving
approximation using \e{cv_def}, \e{mean_lif_small_lam}, and
\e{var_lif_small_lam}.}
\end{figure}

Since the minima in mean and variance vs $\lambda$ are attained at
distinct values of the decay rate, I can expect a complicated behavior
for the relative variability of the FPT. Indeed, the CV as a function
of decay rate shown in \bi{cv_lif} first goes through a minimum
($\lambda_{min}\approx 0.66$), reaches a maximum at a finite decay
rate ($\lambda_{max}\approx 2.8$), and saturates at a CV that is
higher than in the unperturbed case. Compared to the CV of the linear
potential system (cf. \bi{cv_pif}), the decrease at small $\lambda$ is
weaker; furthermore, there is no maximum for the linear system but
only a saturation at large decay rate.\\
For the chosen parameters the FPT from $x=0$ to $x=x_E$ can be
split into two independent FPTs as $T=T_1+T_2$ with $T_1$ being the
FPT from $x=0$ into the minimum $x=a/b$ and $T_2$ being the time for
the passage from the minimum to the absorbing boundary (see, for
instance, ref.~\cite{PakTan01}). It is straightforward to show that
\begin{equation}
\label{cv_split}
CV^2=CV_1 \frac{\lr{T_1}^2}{\lr{T}^2}+CV_2 \frac{\lr{T_2}^2}{\lr{T}^2}
\end{equation}
where $CV_{1}$ and $CV_2$ are the CV of the respective passage
processes. Now the relaxation into the minimum is evidently more
regular than the noise-assisted escape out of the potential minimum,
i.e. $CV_2>CV_1$. The behavior of the CV can be understood by considering
how in the limits of small and large decay rate the relative contributions
of $T_1$ and $T_2$ are changed.\\
At low decay rate, the driving is effectively static, hence the system
is equivalent to the unperturbed dynamics with enlarged bias
$a+\varepsilon\lambda$. This system in turn is equivalent to the
unperturbed dynamics with the original bias $a$ but initial point at
$x=-\varepsilon\lambda$ and absorbing boundary at
$x=x_E-\varepsilon\lambda$. With these parameters, the FPT $T_1$ from
initial point to potential minimum increases and the FPT $T_2$ from
minimum to absorbing boundary drops compared to the unperturbed
case. It is reasonable that the CV of the single passage processes
change only little; what mainly changes is their relative contribution
to the total CV by means of the squared ratios $T_1/T$ and
$T_2/T$. Hence, according to \e{cv_split} the CV will drop since the
lower CV of $T_1$ makes a larger relative contribution. From this line
of argument, it is also reasonable that the CV drops with increasing
$\lambda$ as long as the static approximation holds true.\\
In the limit of $\lambda\gg\langle T\rangle$, one obtains also the
unperturbed dynamics (as explained above by means of
\e{langevin_appro}) but for a passage process from
$x=\varepsilon[1+b/(\lambda-b)]$ to $x=x_E$. Obviously, here the
escape time $T_2$ is the same as in the unperturbed case; the first
FPT $T_1$, however, has been shortened.  According to \e{cv_split},
one can thus expect a {\em higher} CV than in the unperturbed case
since $CV_2$ makes a larger relative contribution to the CV of the
total FPT.  Moreover, the $\lambda\to\infty$ limit of the CV is
approached from above because the shift in initial point drops with
increasing $\lambda$. Interpolating in the simplest way between the
behavior at small and large $\lambda$ predicts a minimum at moderately
low decay rate and a maximum at larger rate as has been found in
\bi{cv_lif}. I note that for the decay rate at which the variance
attains a minimum, the CV is higher than in the unperturbed case,
similarly to the linear potential case ($b=0$) discussed in the
previous subsection.\\
Like in case of the biased random walk, the behavior of mean and
variance as functions of the decay rate depends crucially on the
constant-intensity scaling of the driving I have used. Additionally,
the state-dependence of the force leads to a nonmonotonous behavior of
the mean as a function of the decay rate.  The minima in mean and
variance are not true resonances as in case of a periodic driving but
are mainly related to the distinct sensitivity of the FPT statistics
with respect to changes in the initial point or in the bias parameter,
respectively. Nevertheless, in physical situations where a
constant-intensity scaling of the driving is appropriate, the
nonmonotonous behavior of the first two cumulants and of the CV may be
of some importance.
\section{Summary and outlook}
\label{sec:summary}
In this paper I have studied the moments of the first passage time in
presence of a weak time-dependent driving. A formula for the
corrections to the moments for arbitrary driving and potential shape
were derived that contains, however, the time-dependent probability
density of the unperturbed system or its Laplace transform. The latter
functions are known only in a few rare cases. Explicite correction
formulas for mean and variance of the first passage time could be
achieved for the case of an exponentially decaying driving function
and an either linear or parabolic potential.  These analytical results
were found to be in excellent agreement with results from computer
simulations of the passage processes. I demonstrated furthermore, that
for the chosen exponential driving, the variance of the passage time
in the linear case as well as both the mean and the passage time in
case of a parabolic potential pass through minima as functions of the
decay rate of the driving. The behavior of the relative standard
deviation (i.e. the CV) proved to be even more complicated. All of
these findings resemble the effects of coherent stochastic resonance
and resonant activation, however, they rely on a different mechanism
involving the constant-intensity scaling of the driving.\\
The explicit results for the case of a linear or a parabolic potential
derived in this paper will be applied in the near future to the
neurobiological problems mentioned in the introduction. Furthermore,
the results can be useful for other problems, too. The application of
the formulas to the case of a periodic driving with a cosine function
studied in Refs. \cite{BulLow94,GitWei95,BulEls96} is
straightforward. Likewise, the case of an exponentially damped cosine
function involving two time scales (driving period and decay rate) can
be readily derived from my formulas and might be worth to look at.\\
The general approach presented in this paper may be easily extended to
the cases of two absorbing boundaries or one absorbing and one
reflecting boundary. I am also convinced that the problem of a
state-dependent driving (i.e. dealing with a force $s(x,t)$ instead of
$s(t)$) can be successfully treated with the approach. Finally, the
case of a stochastic driving function might be tackled by a proper
average of the correction formulas over the driving process and its
initial condition. This last problem, though, seems to be much more
challenging than the other extensions of the theory.

\section{Acknowledgment}
\label{sec:acknowledgement}
I am indebted to Andr{\'e} Longtin for inspiring discussions that
brought the subject of this paper to my attention; I furthermore wish
him to thank for his generous support during the last years. This work
has been supported by NSERC Canada.
\appendix
\section{Equivalence of the different  quadratures expressions}
Here I show that \e{jn_0} together with \e{j0_0} and \e{rel_Tn_Jn}
yields the same FPT moments as the standard formula \e{quadratures}.
The two differing expressions for the $n$-th moment can be written as
follows (indices ``S'' and ``A'' stand for ``standard'' and
``alternative'')
\begin{eqnarray}
\lr{T^n}_S&=&\frac{n}{D^n} \!\!\!\int\limits_a^b \!\!dy_1 e^{\frac{U(y_1)}{D}}\!\!\!\int\limits_{-\infty}^{y_1} \!\!dx_1 e^{-\frac{U(x_1)}{D}}\!\!\!\int\limits_{x_1}^b \!\!dy_2 e^{\frac{U(y_2)}{D}}\!\!\!\int\limits_{-\infty}^{y_2} \!\!dx_2 e^{-\frac{U(x_2)}{D}} \cdots\nonumber \\
&& \hspace*{3em}\cdots\int\limits_{x_{n-1}}^b \!\!dy_n e^{\frac{U(y_n)}{D}}\!\!\!\int\limits_{-\infty}^{y_n} \!\!dx_n e^{-\frac{U(x_n)}{D}}\\
\lr{T^n}_A&=&\frac{n}{D^n} \!\!\!\int\limits_{-\infty}^b \!\!dy_1 e^{-\frac{U(y_1)}{D}}\!\!\!\int\limits_{y_1}^{b} \!\!dx_1 e^{\frac{U(x_1)}{D}}\!\!\!\int\limits_{-\infty}^{x_1} \!\!dy_2 e^{-\frac{U(y_2)}{D}}\!\!\!\int\limits_{y_2}^{b} \!\!dx_2 e^{\frac{U(x_2)}{D}} \cdots \nonumber\\
&& \hspace*{3em}\cdots\int\limits_{-\infty}^{x_{n-1}} \!\!dy_n e^{-\frac{U(y_n)}{D}}\!\!\!\int\limits_{y_n}^{b} \!\!dx_n e^{\frac{U(x_n)}{D}}\Theta(x_n-a)
\end{eqnarray}
Note that for the second formula, I have used an arbitrary initial
point $x(t=0)=a$ which only changes the argument of the Heaviside
function in \e{j0_0}.\\
Now I introduce the operators
\begin{eqnarray}   
{\hat{\cal K}}_{x_j}(x)&=& \int\limits_{x_j}^b dy\; e^{U(y)/D} \int\limits_{-\infty}^{y} dx e^{-U(x)/D}\\
{\hat{\cal M}}_{x_j}(x)&=& \int\limits_{-\infty}^{x_j} dy\; e^{-U(y)/D} \int\limits_{y}^{b} dx e^{U(x)/D}
\end{eqnarray}
where the argument indicates the variable with respect to which the
respective function is integrated, while the index denotes as a
parameter one boundary of integration. It is not hard to show that for $u,w<b$
\begin{equation}
\label{rel_ij_theta}
{\hat{\cal M}}_{u}(v) \Theta(v-w)={\hat{\cal K}}_{w}(v)\Theta(u-v). 
\end{equation}
and
\begin{equation}
\label{rel_ij_comm}
{\hat{\cal M}}_{u}(v) {\hat{\cal K}}_{w}(x) ={\hat{\cal K}}_{w}(x){\hat{\cal M}}_{u}(v),
\end{equation} 
i.e., the operators commute if their arguments and indices differ.\\
By means of the operators, the two expressions for the $n$-th moment
can be written as follows
\begin{eqnarray}
\label{tns_op}
\lr{T^n}_S&=&\frac{n}{D^n} {\hat{\cal K}}_a(x_1){\hat{\cal K}}_{x_1}(x_2)\cdots{\hat{\cal K}}_{x_{n-1}}(x_n) \Theta(b-x_n) \\ 
\label{tna_op}
\lr{T^n}_A&=&\frac{n}{D^n} {\hat{\cal M}}_b(x_1){\hat{\cal M}}_{x_1}(x_2)\cdots{\hat{\cal M}}_{x_{n-1}}(x_n) \Theta(x_n-a)
\end{eqnarray}
Note that the function $\Theta(b-x_n)$ in \e{tns_op} is always one
since $x_j<b$ for all $j=1,\dots n$. Using the relations
\e{rel_ij_theta} and \e{rel_ij_comm} it follows from \e{tna_op} that
\begin{eqnarray}
\lr{T^n}_A&=&\frac{n}{D^n} {\hat{\cal M}}_b(x_1)\cdots{\hat{\cal M}}_{x_{n-2}}(x_{n-1}){\hat{\cal K}}_{a}(x_{n}) \Theta(x_{n-1}-x_n),\nonumber \\
&=&\frac{n}{D^n} {\hat{\cal M}}_b(x_1)\cdots{\hat{\cal M}}_{x_{n-3}}(x_{n-2}){\hat{\cal K}}_{a}(x_{n}) {\hat{\cal K}}_{x_n}(x_{n-1}) \Theta(x_{n-2}-x_{n-1}),\nonumber \\
&& \frac{}{}\vdots ,\nonumber \\
&=&\frac{n}{D^n} {\hat{\cal K}}_{a}(x_{n})\cdots{\hat{\cal K}}_{x_2}(x_1)  \Theta(b-x_1)\nonumber \\
&=& \lr{T^n}_S
\end{eqnarray}  
as it should be.
\bibliographystyle{unsrt} 
\bibliography{../../BIB/ALL} \end{document}